# Entropy vectors and network codes

Terence Chan and Alex Grant
Institute for Telecommunications Research
University of South Australia, Australia
{terence.chan, alex.grant}@unisa.edu.au

*Abstract*— We consider a network multicast example that relates the solvability of the multicast problem with the existence of an entropy function. As a result, we provide an alternative approach to the proving of the insufficiency of linear (and abelian) network codes and demonstrate the utility of non-Shannon inequalities to tighten outer bounds on network coding capacity regions.

## I. INTRODUCTION

Traditionally, intermediate nodes in a communication network duplicate and forward packets towards their final destination. Although such a store-and-forward scheme is simple to implement, it does not guarantee efficient utilization of available transmission bandwidth. The network coding approach [1], [2] generalizes the traditional scheme by allowing intermediate nodes to forward mixed/encoded versions of all received data packets. Not only can this significantly increase network throughput in multicast scenarios, it can also provide robustness to link failure, and minimize cost of transmission.

One fundamental problem in network coding is to understand the capacity region and the classes of codes that achieve capacity. In the single session multicast scenario, the problem is well understood. In particular, the capacity region is characterized by max-flow/min-cut bounds and linear network codes are sufficient to achieve maximal throughput [2], [3].

However, things are much more complicated in more general multicast scenarios. It was recently proved that linear network codes are not sufficient for general multicast scenarios [3]. Furthermore, the capacity region achieved by network codes are also unknown. In fact, there are limited tools available to study the capacity region.

One powerful tool is developed in [4], which outerbounds the capacity region by characterizing the intersection of a set of hyperplanes and the set of entropy functions. Unfortunately, it cannot be implemented in practice due to the lack of an explicit characterization of entropy functions for more than three random variables. One way to resolve this difficulty is by replacing the set of entropy functions with the set of non-negative submodular functions. The resulting bounds for the network coding capacity region can be quite loose. Recent work [5] based on matroid theory showed that application of the non-Shannon inequality [6], one can obtain a tighter bound for the capacity region (by obtaining a better outer bound for the set of entropy functions).

In this paper, we construct a particular network and multicast problem, which relates solvability to the existence of an entropy function on a set of four random variables. As a result, we provide an alternative approach to the proving of the insufficiency of linear (and abelian group) network codes, including time-sharing of such network codes. We also demonstrate the utility of non-Shannon inequalities to tighten outer bounds on network coding capacity regions.

Section II provides (linear/group) network code fundamentals and relates the existence of such codes to the ranks of certain vector subspaces/group orders. Section III presents the main result, concerning a multicast problem induced by a function $h$ with certain properties. Theorem 1 states that $h$ is entropic if the induced multicast problem is solvable. Theorem 2 provides a converse, namely that the multicast problem is solvable if $h$ is entropic. Section IV explores the implications of this result, which include the insufficiency of linear or even (abelian) group network codes, and the neccessity for non-Shannon inequalities.

## II. NETWORKS, CODES AND CAPACITY REGIONS

A communication network will be modeled by a directed acyclic graph $\mathcal{G} = (\mathcal{N}, \mathcal{E})$. The nodes $u \in \mathcal{N}$ and directed edges $e = (\text{tail}(e), \text{head}(e)) \in \mathcal{E}$ respectively model the communication nodes and directed point-to-point communication links in the network. For $e, f \in \mathcal{E}$ we write $f \to e$ as shorthand for $\text{head}(f) = \text{tail}(e)$. Similarly, for $f \in \mathcal{E}, u \in \mathcal{N}$, $f \to u$ means $\text{head}(f) = u$ and $u \to f$ means $\text{tail}(f) = u$.

For any network $\mathcal{G}$, a *multicast requirement* $M \triangleq (\mathcal{S}, O, D)$ is specified by:

1) An index set $\mathcal{S}$ of all independent multicast sessions. Each session is a collection of data packets to be multicast to a prescribed set of destination nodes.
2) A mapping $O : \mathcal{S} \mapsto \mathcal{N}$, where $O(s)$ is the originating node for multicast session $s$.
3) A mapping $D : \mathcal{S} \to 2^{\mathcal{N}}$, where the set $D(s)$ contains the destination nodes for the session $s$, all of which require the data of session $s$.

*Note: We do not specify any rate requirement.* Given a multicast requirement $M$, the objective is to efficiently multicast data for session $s$ originating at $O(s)$ to all destinations $D(s)$.

For a given network $\mathcal{G}$ and multicast requirement $M$, a network code solving the multicast problem is specified by a set of alphabets $\mathcal{V} = \{\mathcal{V}_f, f \in \mathcal{S} \cup \mathcal{E}\}$ and a set of local coding functions $\Phi \triangleq \{\phi_e : e \in \mathcal{E}\}$.

Data transmission takes place as follows. The input generated by session $s$ is a symbol $V_s \in \mathcal{V}_s$. The symbol transmitted along edge $e \in \mathcal{E}$ is

$$V_e = \phi_e(V_f : f \in \mathcal{S} \cup \mathcal{E}, f \to e),$$

where to simplify notation, $s \to e$ indicates $O(s) \to e$. We require that for every $s \in \mathcal{S}$ and $u \in D(s)$, $V_s$ can be uniquely determined from $\{V_f : f \to u\}$

$$H(V_s \mid V_f : f \in \mathcal{S} \cup \mathcal{E}, f \to u) = 0.$$

We will refer to a network code by $\Phi$, with the alphabet $\mathcal{V}$ implicitly defined. Since the source, and hence edge symbols $V_f$ are random variables, we can also refer to the code by the collection of random variables $\{V_f, f \in \mathcal{S} \cup \mathcal{E}\}$, where their joint distribution is implied by $\Phi$.

**Definition 1.** *Given a network $\mathcal{G}$ and a multicast requirement $M$, a rate-capacity tuple*

$$(\lambda, \omega) \triangleq (\lambda_s : s \in \mathcal{S}, \omega_e : e \in \mathcal{E})$$

*is* admissible *if there exists a network code $\Phi$ such that*

$$\log |\mathcal{V}_e| \leq \omega_e, \forall e \in \mathcal{E}$$
$$\log |\mathcal{V}_s| \geq \lambda_s, \forall s \in \mathcal{S}$$

*Similarly, a tuple is* asymptotically admissible *if there exists a sequence of network codes $\Phi^{(n)}$ over alphabets $\mathcal{V}^{(n)}$ and positive normalizing constants $r(n)$ such that*

$$\limsup_{n \to \infty} \frac{1}{r(n)} \log |\mathcal{V}_e^{(n)}| \leq \omega_e, \forall e \in \mathcal{E},$$
$$\liminf_{n \to \infty} \frac{1}{r(n)} \log |\mathcal{V}_s^{(n)}| \geq \lambda_s, \forall s \in \mathcal{S}$$

*Furthermore, if the sequence of codes all belong to a particular class (e.g. linear) we call the rate-capacity tuple asymptotically admissible by that class of codes.*

For a given network, multicast requirement and rate-capacity tuple, the multicast problem is to determine if the tuple is (asymptotically) admissible. Equivalently determine the set of all (asymptotically) admissible rate-capacity tuples.

In the above formulation, there is no restriction on the choice of alphabets and local coding functions. However, network codes with neat algebraic structures may be preferred in practice, to reduce encoding and decoding complexity.

**Definition 2.** *A network code $\Phi$ is* linear *over $\mathbb{F}_q$ if $\mathcal{V}_f$ is a vector space over $\mathbb{F}_q$, and the $\phi_e$ are linear.*

**Proposition 1.** *Suppose $\Phi$ is linear. Then there exists a vector space $\mathbf{V}$ with subspaces $\mathbf{V}_f, f \in \mathcal{E} \cup \mathcal{S}$ such that*

$$|\mathcal{S}| \dim \mathbf{V} - \sum_{s \in \mathcal{S}} \dim \mathbf{V}_s = \dim \mathbf{V} - \dim \bigcap_{s \in \mathcal{S}} \mathbf{V}_s$$

$$\bigcap_{f:f \to e} \mathbf{V}_f \text{ is a subspace of } \mathbf{V}_e, \forall e \in \mathcal{E}$$

$$\bigcap_{f:f \to u} \mathbf{V}_f \subseteq \mathbf{V}_s, \forall s \in \mathcal{S}, u \in D(s)$$

$$\dim \mathbf{V} - \dim \mathbf{V}_s = \log_q |V_s|, \forall s \in \mathcal{S}$$
$$\dim \mathbf{V} - \dim \mathbf{V}_f \leq \log_q |V_f|, \forall f \in \mathcal{E}$$

**Proposition 2.** *Suppose that there exists a vector space $\mathbf{V}$ with subspaces $\mathbf{V}_f, f \in \mathcal{E} \cup \mathcal{S}$ satisfying*

$$|\mathcal{S}| \dim \mathbf{V} - \sum_{s \in \mathcal{S}} \dim \mathbf{V}_s = \dim \mathbf{V} - \dim \bigcap_{s \in \mathcal{S}} \mathbf{V}_s$$

$$\bigcap_{f:f \to e} \mathbf{V}_f \text{ is a subgroup of } \mathbf{V}_e, \forall e \in \mathcal{E}$$

$$\bigcap_{f:f \to u} \mathbf{V}_f \subseteq \mathbf{V}_s, \forall s \in \mathcal{S}, u \in D(s)$$

*Then there exists a linear network code $\Phi$ such that $\log_q |\mathcal{V}_f| = \dim \mathbf{V} - \dim \mathbf{V}_f$ for all $f \in \mathcal{S} \cup \mathcal{E}$. In other words, the rate-capacity tuple $(\lambda_s : s \in \mathcal{S}, \omega_e : e \in \mathcal{E})$ where $\lambda_s = (\dim \mathbf{V} - \dim \mathbf{V}_s) \log q$ and $\omega_e = (\dim \mathbf{V} - \dim \mathbf{V}_e) \log q$ is linearly admissible.*

The above two propositions show that linear network codes can be described by the ranks of a collection of vector subspaces. This idea can be extended to other algebraic structures.

**Proposition 3.** *Let $G$ be a finite group with subgroups $G_f, f \in \mathcal{S} \cup \mathcal{E}$. Suppose the following conditions are satisfied*

$$|G|/|\bigcap_{s \in \mathcal{S}} G_s| = \prod_{s \in \mathcal{S}} |G|/|G_s|$$

$$\bigcap_{f:f \to e} G_f \text{ is a subgroup of } G_e, \forall e \in \mathcal{E}$$

$$\bigcap_{f:f \to u} G_f \text{ is a subgroup of } G_s, \forall s \in \mathcal{S}, u \in D(s),$$

*Then there exists a network code such that the rate-capacity tuple $(\log |G|/|G_s| : s \in \mathcal{S}, \log |G|/|G_e| : e \in \mathcal{E})$ is admissible.*

If the group $G$ is also abelian, the group network code will be called abelian.

## III. MAIN RESULT

Let $h$ be a set function defined on the collection of all nonempty subsets of $\{1, 2, 3, 4\}$ such that

$$h(1,2,3,4) = h(1,2,3) = h(1,2,4) =$$
$$h(1,3,4) = h(2,3,4) = h(3,4). \quad (1)$$

Then $h$ induces a multicast problem $\text{MP}(h)$, together with an induced rate-capacity tuple, which we will now describe. There are three sessions $V_a, V_b, V_c$ all available at the same source node, with entropy rates $H(V_a) = h(1)$, $H(V_b) = h(1,2) - h(1)$ and $H(V_c) = h(1,2,3) - h(1,2)$.

For convenience, we divide the underlying network into two parts. The first part, depicted in Figure 1, contains the source and generates a set of four network coded messages $V_1, V_2, V_3, V_4$ which will be transmitted to the second half of the network, shown in Figure 2. To simplify notation, each edge is associated with a symbol denoting the network code message (and corresponding random variable) to be transmitted along that edge. If the link is capacitated, its capacity is given following the edge message symbol.

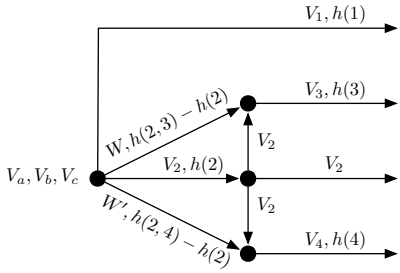

Fig. 1. First part of the network

The second part of the network is mainly composed of receivers, and is divided into several subnetworks, Figures 2(a)–2(e) each of which contains a receiver and a subset of $V_1, V_2, V_3, V_4$ as input. Receivers are denoted by a solid square labeled by the set of source messages to be recovered. To simplify notation, all messages other than the four messages $V_1, V_2, V_3, V_4$ will be denoted by a generic symbol $W$. As each receiver subnetwork has at most one such message, any confusion will be cleared from the context.

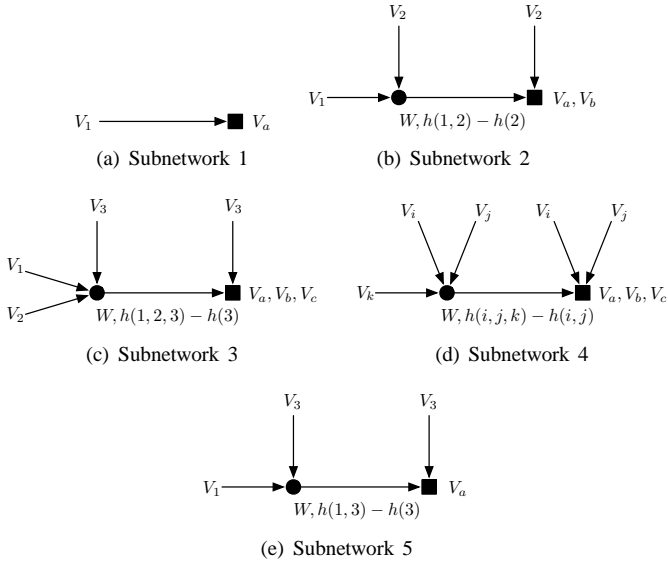

Fig. 2. Second part of the network

The multicast problem that we are interested in is to determine if there is any network coding solution for the induced multicast problem. Equivalently, we wish to determine if the multicast problem $\text{MP}(h)$ is (asymptotically) solvable or not.

**Theorem 1.** *Let $h$ be a set function defined on the collection of all nonempty subsets of $\{1, 2, 3, 4\}$ satisfying (1). If the induced multicast problem $\text{MP}(h)$ is solvable, then $h$ is the entropy vector for the random variables $V_1, V_2, V_3, V_4$ (see Figure 1 and 2) of the network coding solution. Hence, $h$ is entropic.*

*Proof.* Suppose $\{V_a, V_b, V_c, V_1, V_2, V_3, V_4\}$ is a solution to the multicast problem $\text{MP}(h)$. Then $H(V_a) = h(1)$, $H(V_b) = h(1,2) - h(1)$ and $H(V_c) = h(1,2,3) - h(1,2)$.

From the first half of the network, it is obvious that $H(V_i) \leq h(i)$ for $i = 1, 2, 3, 4$. Also, we can deduce

$$\begin{aligned} H(V_2, V_3) &\leq H(V_2, W, V_3) \\ &= H(V_2, W) \\ &\leq H(V_2) + H(W) \\ &\leq h(2) + h(2,3) - h(2). \end{aligned}$$

Similarly, $H(V_2, V_4) \leq h(2, 4)$.

To prove that $h$ is the entropy vector for $V_1, V_2, V_3, V_4$, we need to show that $H(V_i, i \in \alpha) = h(\alpha)$ for all $\alpha \subseteq \{1, 2, 3, 4\}$. This is accomplished by Claims 1–9 below. □

**Claim 1.** $H(V_1) = h(1)$ and $H(V_1|V_a) = 0$

*Proof.* From Figure 2(a), $H(V_a|V_1) = 0$ in order to recover $V_a$ at the sink. Therefore, $H(V_1) \geq H(V_a) = h(1)$. As $H(V_1) \leq h(1)$, the first part of the claim follows. Moreover,

$$\begin{aligned} H(V_1|V_a) &= H(V_1, V_a) - H(V_a) \\ &= H(V_1) + H(V_a|V_1) - H(V_a) \\ &= H(V_1) - H(V_a) \\ &= 0. \end{aligned}$$
□

**Claim 2.** $H(V_2) = h(2)$ and $H(V_1, V_2) = h(1, 2)$

*Proof.* From Figure 2(b),

$$\begin{aligned} H(V_2) + h(1,2) - h(2) &\geq H(V_2) + H(W) \\ &\geq H(V_2, W) \\ &\geq H(V_a, V_b) \\ &= H(V_a) + H(V_b) \\ &= h(1,2) \end{aligned}$$

As a result, $H(V_2) \geq h(2)$. Since $H(V_2) \leq h(2)$, we have $H(V_2) = h(2)$. On the other hand, from Figure 2(b), $H(V_1, V_2) \geq H(V_a, V_b) = h(1, 2)$. Finally, note that

$$\begin{aligned} H(V_1, V_2) &\leq H(V_1, V_2, V_a, V_b, W) \\ &\leq H(V_2, W, V_a, V_b) + H(V_1|V_a) \\ &= H(V_2, W) + H(V_a, V_b|V_2, W) \\ &\leq H(V_2) + H(W) \\ &= h(2) + h(1,2) - h(2) \\ &= h(1,2). \end{aligned}$$
□

**Claim 3.** $H(V_3) = h(3)$ and $H(V_4) = h(4)$

*Proof.* From Figure 2(c),

$$\begin{aligned} H(V_3) + h(1,2,3) - h(3) &\geq H(V_3) + H(W) \\ &\geq H(V_a, V_b, V_c) \\ &= h(1,2,3) \end{aligned}$$

Hence $H(V_3) \geq h(3)$ and since $H(V_3) \leq h(3)$, $H(V_3) = h(3)$. Similarly, $H(V_4) = h(4)$. □

**Claim 4.** $H(V_i, V_j) \geq h(i,j)$ for distinct $i$ and $j$

*Proof.* From Figure 2(d),

$$H(V_i, V_j) + h(i,j,k) - h(i,j) \geq H(V_i, V_j) + H(W)$$
$$\geq H(V_a, V_b, V_c)$$
$$= h(i,j,k)$$

As a result, $H(V_i, V_j) \geq h(i,j)$. □

**Claim 5.** $H(V_1, V_3) = h(1,3)$ and $H(V_1, V_4) = h(1,4)$.

*Proof.* From Figure 2(e),

$$H(V_1, V_3) \leq H(V_1, V_3, W, V_a)$$
$$\leq H(V_3, W) + H(V_a|W, V_3) + H(V_1|V_a)$$
$$\leq H(V_3) + H(W)$$
$$\leq h(3) + h(1,3) - h(3)$$
$$= h(1,3)$$

Thus $H(V_1, V_3) = h(1,3)$. Similarly, $H(V_1, V_4) = h(1,4)$. □

**Claim 6.** $H(V_i, V_j, V_k) = h(i,j,k)$

*Proof.* From Figure 2(d), it is clear that $H(V_i, V_j, V_k) \geq H(V_a, V_b, V_c)$. On the other hand, as $V_i, V_j, V_k$ are functions of $V_a, V_b, V_c$, therefore, $H(V_i, V_j, V_k) \leq H(V_a, V_b, V_c)$ and hence $H(V_i, V_j, V_k) = H(V_a, V_b, V_c) = h(i,j,k)$. □

**Claim 7.** $H(V_3, V_4) = h(3,4)$.

*Proof.* From Claim 4, we have $H(V_3, V_4) \geq h(3,4)$. Since $h(3,4) = H(V_a, V_b, V_c)$ and $H(V_3, V_4|V_a, V_b, V_c) = 0$, the claim then follows. □

**Claim 8.** $H(V_1, V_2, V_3, V_4) = H(V_a, V_b, V_c) = h(1,2,3,4)$

*Proof.* Obvious □

**Claim 9.** $H(V_2, V_3) = h(2,3)$ and $H(V_2, V_4) = h(2,4)$

*Proof.* From Claim 4, $H(V_2, V_3) \geq h(2,3)$ and $H(V_2, V_4) \geq h(2,4)$. As $H(V_2, V_3) \leq h(2,3)$ and $H(V_2, V_4) \leq h(2,4)$ are proved in the beginning, the result then follows. □

**Corollary 1.** *If the induced multicast problem $\text{MP}(h)$ is asymptotically solvable, then $h \in \overline{\Gamma_4^*}$ which is the minimal closure of the set of all entropy functions for four random variables.*

Theorem 1 shows that if the induced multicast problem $\text{MP}(h)$ is solvable, then $h$ must be entropic. In the following theorem, we prove the converse.

**Theorem 2.** *Let $h$ be entropic and that all the terms $h(1,2,3,4), h(1,2,3), h(1,2,4), h(1,3,4), h(2,3,4), h(3,4)$ are equal to each other. Then the rate-capacity tuple in the induced multicast problem $\text{MP}(h)$ is asymptotically admissible.*

*Proof.* As $h$ is entropic, then there exists a set of random variables $(U_1, U_2, U_3, U_4)$ whose entropy function is $h$. Hence, one can construct a sequence of groups $G^k$, with subgroups $G_i^k, i \in \{1,2,3,4\}$ [7], [8] and normalizing constants $r(k)$ such that

$$\lim_{k \to \infty} \frac{1}{r(k)} h^k(\alpha) = h(\alpha), \alpha \subseteq \{1,2,3,4\}$$
$$h^k(1,2,3,4) = h^k(1,2,3) = h^k(1,2,4) = h^k(1,3,4)$$
$$= h^k(2,3,4) = h^k(3,4).$$

where $h^k(\alpha) = \log |G^k|/|\cap_{i \in \alpha} G_i^k|$.

For any $k$, the function $h^k$ induces a multicast problem $\text{MP}(h^k)$. If $\text{MP}(h^k)$ is solvable, then the theorem follows from $\lim_{k \to \infty} h^k(\alpha)/r(k) = h(\alpha)$.

It remains to show that $\text{MP}(h^k)$ is solvable. For each $k$, $h^k$ is entropic [7], [8]. In fact, $\{G^k, G_i^k : i \in \{1,2,3,4\}\}$ induces a set of *quasi-uniform* random variables $U_1^k, U_2^k, U_3^k, U_4^k$ such that

1) For any $\alpha \subseteq \{1,2,3,4\}$, $\{U_i^k : i \in \alpha\}$ is uniformly distributed over its support.
2) For any $\alpha, \beta \subseteq \{1,2,3,4\}$, the conditional probability distribution of $\{U_i^k : i \in \alpha\}$ given a particular instance of $\{U_i^k : i \in \beta\} = \{u_i^k : i \in \beta\}$ (with positive probability) is constant over its support.

By the quasi-uniformity of $U_1^k, U_2^k, U_3^k, U_4^k$, it is straightforward to show that there exists another set of quasi-uniform random variables $U_a^k, U_b^k, U_c^k$ such that $U_a^k, U_b^k, U_c^k$ are independent and all the terms $H(U_1^k|U_a^k)$, $H(U_a^k|U_1^k)$, $H(U_a^k, U_b^k|U_1^k, U_2^k)$, $H(U_1^k, U_2^k|U_a^k, U_b^k)$, $H(U_a^k, U_b^k, U_c^k|U_1^k, U_2^k, U_3^k)$, $H(U_1^k, U_2^k, U_3^k|U_a^k, U_b^k, U_c^k)$ are equal to zero.

It can be checked easily that $U_a^k, U_b^k, U_c^k$ are uniform and that $H(U_a^k) = h^k(1)$, $H(U_b^k) = h^k(1,2) - h^k(1)$ and $H(U_c^k) = h^k(1,2,3) - h^k(1,2)$. To show that $\text{MP}(h^k)$ is solvable, it remains to show that the auxiliary random variables (i.e., the generic message random variables $W$) can be constructed to satisfy the capacity constraints.

This can be achieved by "data compression" of the inputs subject to common side-information at both encoder and decoder. Since $U_1^k, U_2^k, U_3^k, U_4^k$ are quasi-uniform, it is straightforward to verify that the "rate of compression" meets the (conditional) entropy lower bounds by using block coding. The result then follows. □

## IV. IMPLICATIONS

**Theorem 3.** *There is a network and a multicast requirement such that the use of abelian network codes is suboptimal (including linear network codes, R–module codes, and time-sharing of such).*

*Proof.* First, we describe a set of four quasi-uniform random variables $U_1, U_2, U_3, U_4$ constructed using the projective plane described in [6]. The joint entropies of subsets of these random

variables are as follows:

$$h(1) = h(2) = h(3) = (4) = \log 13$$
$$h(1,2) = \log 6 + \log 13; h(3,4) = \log 13 + \log 12$$
$$h(1,3) = h(1,4) = h(2,3) = h(2,4) = \log 13 + \log 4$$
$$h(i,j,k) = \log 13 + \log 12 = h(1,2,3,4), \ \forall \text{ distinct } i,j,k.$$

Then it is clear that $h(1,2,3,4) = h(1,2,3) = h(1,2,4) = h(1,3,4) = h(2,3,4) = h(3,4)$. Since $h$ is entropic, $\text{MP}(h)$ is asymptotically solvable. In fact, it can be shown that $h$ is an entropy vector of four quasi-uniform random variables. Therefore, using the same argument as used in the proof of Theorem 2, we can show that $\text{MP}(h)$ is solvable.

Suppose to the contrary that there is an abelian network code that solves the multicast problem. Then there exists an abelian group $G$ with subgroups $G_1, G_2, G_3, G_4$ such that $h(\alpha) = \log |G|/|\bigcap_{i \in \alpha} G_i|$. By [8], $h$ must satisfy the Ingleton inequality

$$h(1,2) + h(1,3) + h(1,4) + h(2,3) + h(2,4) \geq$$
$$h(1) + h(2) + h(3,4) + h(1,2,3) + h(1,2,4) \quad (2)$$

However, we can directly verify that $h$ does not satisfy the Ingleton inequality. □

**Corollary 2.** *There is a network and a multicast requirement for which abelian codes are (asymptotically) suboptimal.*

The multicast network constructed can not only be used to show the suboptimality of abelian network codes, it also shows that non-Shannon type information inequalities can be useful in determining the capacity region of network codes.

Suppose $h(1,2,3,4) = h(1,2,3) = h(1,2,4) = h(1,3,4) = h(2,3,4) = h(3,4)$. By Theorems 1 and 2, we proved that the multicast problem $\text{MP}(h)$ is asymptotically solvable if and only if $h \in \overline{\Gamma_4^*}$. Therefore, we can obtain an outer bound for network code capacity region by finding an outer bound for $\overline{\Gamma_4^*}$. One simple outer bound is $\Gamma_4$ defined as the set of all non-negative submodular functions.

It was shown in [6] that $\Gamma_4$ is a proper superset of $\overline{\Gamma_4^*}$. Therefore, by Theorem 2, the outer bound for the capacity region obtained in this manner is not tight. To tighten the bound for the capacity region, one approach is to tighten the bound for $\overline{\Gamma_4^*}$ by applying new non-Shannon inequalities [6] and in [9].

Consider the function $h$ defined as follows [6]:

$$h(1) = h(2) = h(3) = (4) = 2a > 0$$
$$h(1,2) = 3a; h(3,4) = 4a$$
$$h(1,3) = h(1,4) = h(2,3) = h(2,4) = 3a$$
$$h(i,j,k) = 4a = h(1,2,3,4), \ \forall \text{ distinct } i,j,k.$$

Clearly, $h(1,2,3,4) = h(1,2,3) = h(1,2,4) = h(1,3,4) = h(2,3,4) = h(3,4)$. It can be verified directly that $h \in \Gamma_4$. However, the non-Shannon information inequality obtained in [6] shows that $h \notin \overline{\Gamma_4^*}$. Therefore, the multicast problem $\text{MP}(h)$ is not asymptotically solvable.

## V. CONCLUSION

We have shown how to construct a multicast problem from a function $h$ that is solvable if and only if $h$ is entropic. This provides a useful link between entropy vectors and the capacity region for network codes. Using this approach, we provide an alternative proof of the insufficiency of linear (and abelian group) network codes, including time-sharing of such network codes. We also demonstrated the utility of the Zhang-Yeung inequality to tighten outer bounds on network coding capacity regions.

## ACKNOWLEGEMENT

This work was supported by the Australian Government under ARC grant DP0557310, and by the Defence Science and Technology Organisation under the contract numbers 4500485167 and 4500550654.